\documentclass[sigconf, screen]{acmart}
\AtBeginDocument{%
  \providecommand\BibTeX{{%
    \normalfont B\kern-0.5em{\scshape i\kern-0.25em b}\kern-0.8em\TeX}}}

\copyrightyear{2022}
\acmYear{2022}
\setcopyright{acmlicensed}\acmConference[KDD '22]{Proceedings of the 28th ACM SIGKDD Conference on Knowledge Discovery and Data Mining}{August 14--18, 2022}{Washington, DC, USA}
\acmBooktitle{Proceedings of the 28th ACM SIGKDD Conference on Knowledge Discovery and Data Mining (KDD '22), August 14--18, 2022, Washington, DC, USA}
\acmPrice{15.00}
\acmDOI{10.1145/3534678.3539353}
\acmISBN{978-1-4503-9385-0/22/08}

\usepackage{multicol,multirow}
\usepackage{bbding}
\usepackage{subcaption}
\newcommand{\red}[1]{\textcolor{red}{#1}}
\newcommand{\blue}[1]{\textcolor{blue}{#1}}

\DeclareMathOperator*{\argmax}{arg\,max}
\newcommand{\mE}{\mathbb{E}}
\newcommand{\mP}{\mathbb{P}}

\newcommand{\calO}{\mathcal{O}}
\newcommand{\calU}{\mathcal{U}}
\newcommand{\sumU}{\sum_{u\in \calU}}
\newcommand{\calI}{\mathcal{I}}
\newcommand{\sumI}{\sum_{i\in \calI}}
\newcommand{\sumK}{\sum_{k=1}^{n}}

\newcommand{\argmaxI}{\argmax_{\{X_{*,i,*}^{\pi}\}_{i\in \calI}}}

\newcommand{\expo}{\textit{Exp}}
\newcommand{\merit}{\textit{Merit}}
\newcommand{\imp}{\textit{Imp}}
\newcommand{\te}{\textit{TE}}
\newcommand{\allocation}{X_{u,i,k}^{\pi}}
\newcommand{\pimax}{\pi_{\textit{max}}}
\newcommand{\piexpo}{\pi_{\textit{expo-fair\,}}}
\newcommand{\piunif}{\pi_{\textit{unif\,}}}
\newcommand{\pinsw}{\pi_{\textit{NSW}}}
\newcommand{\pinswalpha}{\pi_{\alpha\textit{-NSW}}}
\newcommand{\pinswone}{\pi_{1\textit{-NSW}}}
\newcommand{\reltrue}{r_{\textit{true}}}
\newcommand{\relpred}{r_{\textit{pred}}}

\begin{document}
\title{Fair Ranking as Fair Division:\\Impact-Based Individual Fairness in Ranking}

\author{Yuta Saito}
\affiliation{%
  \institution{Cornell University}
  \city{Ithaca}
  \state{NY}
  \country{USA}
}
\email{ys552@cornell.edu}

\author{Thorsten Joachims}
\affiliation{%
 \institution{Cornell University}
  \city{Ithaca}
  \state{NY}
  \country{USA}
}
\email{tj@cs.cornell.edu}

\renewcommand{\shortauthors}{Saito and Joachims.}

\begin{abstract}
Rankings have become the primary interface in two-sided online markets. Many have noted that the rankings not only affect the satisfaction of the users (e.g., customers, listeners, employers, travelers), but that the position in the ranking allocates exposure -- and thus economic opportunity -- to the ranked items (e.g., articles, products, songs, job seekers, restaurants, hotels). This has raised questions of fairness to the items, and most existing works have addressed fairness by explicitly linking item exposure to item relevance. However, we argue that any particular choice of such a link function may be difficult to defend, and we show that the resulting rankings can still be unfair. To avoid these shortcomings, we develop a new axiomatic approach that is rooted in principles of {\em fair division}. This not only avoids the need to choose a link function, but also more meaningfully quantifies the impact on the items beyond exposure. Our axioms of \textit{envy-freeness} and \textit{dominance over uniform ranking} postulate that for a fair ranking policy every item should prefer their own rank allocation over that of any other item, and that no item should be actively disadvantaged by the rankings. To compute ranking policies that are fair according to these axioms, we propose a new ranking objective related to the \textit{Nash Social Welfare}. We show that the solution has guarantees regarding its envy-freeness, its dominance over uniform rankings for every item, and its Pareto optimality. In contrast, we show that conventional exposure-based fairness can produce large amounts of envy and have a highly disparate impact on the items. Beyond these theoretical results, we illustrate empirically how our framework controls the trade-off between impact-based individual item fairness and user utility.
\end{abstract}


\maketitle

\section{Introduction}
Ranking interfaces are widely used to mediate online market platforms, ranging from booking a hotel to finding qualified employees. 
In these two-sided markets, not only the users (e.g., travelers and employers) obtain utility from the rankings, but the ranking policy also has a crucial impact on the items being ranked (e.g., hotels and job seekers). Many have noted that conventional ranking approaches \citep{robertson1977probability}, which exclusively optimize the utility to the users, may be unfair to the items. In particular, conventional rankings can give items with similar merit (i.e., relevance) very dissimilar exposure and thus economic opportunity~\citep{wang2021fairness,singh2018fairness,biega2018equity,patro2022fair}.

To address this disparity of treatment, existing works on fairness-in-ranking explore how to explicitly link merit to the exposure that is allocated to individual items or groups of items~\citep{biega2018equity,singh2018fairness,singh2019policy,yadav2021policy,yang2021maximizing,morik2020controlling,zehlike2020reducing,do2021two}. Concrete definitions of exposure-based fairness were introduced in the seminal works by \citet{singh2018fairness} and \citet{biega2018equity}, enforcing constraints to allocate exposure proportional to relevance in expectation over groups or amortized over a sequence of rankings. This fairness-of-exposure approach has since been extended to end-to-end policy learning~\citep{singh2019policy,yadav2021policy,oosterhuis2021plrank}, dynamic ranking~\citep{morik2020controlling,yang2021maximizing}, evaluation metrics~\citep{diaz2020evaluating}, and contextual bandits~\citep{wang2021fairness,jeunen2021top}. However, a key criticism of exposure-based fairness is the necessity to select a function that links merit and exposure, and there is typically no justification that this relationship should be linear -- nor would any other choice be more defendable. Furthermore, we show that the fairness-of-exposure approach can still lead to rankings that violate basic fairness principles.

To overcome these shortcomings, we develop an axiomatic approach to fairness in ranking that is rooted in principles of fair division. In particular, we provide a novel formulation of fairness in ranking as a resource allocation problem, where the resources are the positions in a ranking that need to be fairly divided among the items. To concisely define fairness, we propose \textit{envy-freeness} and \textit{dominance over uniform ranking} as two fairness axioms that parallel widely accepted concepts from the fair-division literature~\citep{varian1974equity,foley1966resource,brandt2016handbook}. Envy-freeness requires that no item can gain better impact by exchanging its position allocation with another item in a market. Dominance over uniform ranking requires that every item should gain better impact than under the uniform random ranking policy, ensuring that all items draw some benefit from participating in the platform. This axiomatization not only avoids the need for an arbitrary link function between relevance and exposure, it also directly captures the {\em impact} (e.g. clicks, conversions, revenue, streams) of the ranking on the items, not just their visibility as in the exposure-based framework. Prior attempts to extend fairness-of-exposure to individual fairness of impact have already been shown to lead to degenerate ranking policies~\citep{singh2018fairness}, making ours the first viable formulation of \textit{impact-based individual item fairness}.

The key remaining problem is to find an algorithm for computing rankings that are guaranteed to fulfill the fairness axioms, and that also guarantee a notion of optimality. To this effect, we introduce a new objective for optimizing rankings that is related to the \textit{Nash Social Welfare} (NSW). Under our formulation, maximizing the NSW is a convex program that is efficiently solvable. We prove that the ranking policy that maximizes the NSW-objective has guarantees regarding the envy-freeness of the position allocations and its dominance over the uniform ranking policy for each item. Furthermore, the rankings are Pareto optimal. We also develop an extension of the NSW called the $\alpha$-NSW, which enables a steerable trade-off between user utility and item fairness through a single hyper-parameter. Beyond these conceptual and methodological contributions, we conduct extensive experiments on synthetic and real-world data. We find that conventional exposure-based fairness can produce large amounts of envy and disparate impact on the items. In contrast, the proposed NSW-maximizing policy achieves an almost envy-free impact distribution even with noisy relevance predictions, and provides more equitable benefit to all items.

\section{Related Work}

\paragraph{Fair Ranking in Two-Sided Markets}
There exist several notions of fairness in ranking, ranging from the ones based on the composition of a prefix of the ranking~\citep{asudeh2019designing,celis2017ranking,celis2020interventions,yang2017measuring} to the exposure-based item fairness. The latter argues that fairness in ranking corresponds to how ranking policies allocate exposure to individual items or groups of items based on their merit~\citep{singh2018fairness,singh2019policy,zehlike2020reducing,yadav2021policy,morik2020controlling,yang2021maximizing,singh2021fairness}. Within the exposure-based framework, \citet{singh2018fairness} introduce a post-processing algorithm to allocate exposure for protected item groups proportional to their merit. Following \citet{singh2018fairness}, other fairness-of-exposure methods have been proposed such as end-to-end policy learning~\citep{singh2019policy,yadav2021policy} and dynamic ranking algorithms~\citep{morik2020controlling,yang2021maximizing}. 
Instead of guaranteeing \textit{group} fairness, there is also a line of work aiming at achieving a fair allocation of exposure among \textit{individual} items~\citep{biega2018equity,bower2020individually}. \citet{biega2018equity} solve an integer program and optimize amortized share of exposure over time among each item. \citet{bower2020individually} propose an optimal-transport-based regularizer to enforce individual item fairness.

Although the exposure-based formulation is widely accepted for both group and individual item fairness, there is no obvious choice for how exposure should be linked to merit. Most studies require exposure to be proportional to merit --- but why not proportional to merit squared or some other function? Indeed, no principled justification has been given for any choice of the link function. Our axiomatic approach overcomes this shortcoming, since it does not rely on an arbitrary link function between exposure and merit, but is solely based on the impact of the ranking on the items. Moreover, we show that the existing \textit{exposure-based} framework can create a substantially \textit{disparate impact} among individual items. While a disparate impact constraint has been proposed to ensure \textit{group} item fairness~\citep{singh2018fairness,morik2020controlling}, imposing such constraints per item for individual fairness leads to the uniform random ranking policy~\citep{singh2018fairness}. We are the first to formulate and enforce \textit{impact-based individual item fairness} in a meaningful way.

\paragraph{Fair Division}
\textit{Fair division} has been studied extensively in algorithmic game theory~\citep{brandt2016handbook,foley1966resource,varian1974equity}. The goal of this line of work is to allocate a set of valuable -- but limited -- resources or goods to the agents in a fair manner. The classical fairness desiderata considered in this field are \textit{envy-freeness} (EF) and \textit{proportional fair share} (PFS)~\citep{patro2020fairrec,foley1966resource,brandt2016handbook}. EF requires that no agents prefer another agent’s allocation of goods to their own. PFS means that every agent receives $1/n$ of their utility over the entire set of goods, where $n$ is the number of agents. When the goods are \textit{divisible}, maximizing the \textit{Nash Social Welfare}, the product of agent utilities, has been considered compelling, because its solution ensures EF and PFS as well as being Pareto optimal. In contrast, recent studies focus on a more challenging problem of fairly allocating \textit{indivisible} goods (e.g., course seats in universities~\citep{budish2011combinatorial} or computational resources in a cloud computing environment~\citep{biswas2018fair}). When the goods are indivisible, no feasible allocation may satisfy EF or PFS~\citep{caragiannis2019unreasonable}. Therefore, relaxed versions of these axioms have been considered such as \textit{envy-freeness up to one good} (EF1) and \textit{maximin share guarantee} (MMS)~\citep{brandt2016handbook,patro2020fairrec}. For example, EF1 requires that any pairwise envy can be eliminated by removing a single good from the envied agent’s allocation. An allocation satisfying EF1 always exists for a broad class of utility functions even in the case of indivisible goods~\citep{caragiannis2019unreasonable}.

The notion of EF has been adopted to fair machine learning in binary classification~\citep{balcan2019envy,hossain2020designing}. This is because EF is intuitive and requires no information beyond individuals’ utility functions, which contrasts with other notions of fairness such as metric-based individual fairness~\citep{dwork2012fairness}. Beyond binary classification, a few works have applied the axioms in fair division to recommender system applications. \citet{patro2020fairrec} map the fair recommendation problem in two-sided markets to the problem of fairly allocating indivisible goods, where the goods are the set of products. The goal of \citet{patro2020fairrec} is then to ensure EF1 among users in terms of their utility and MMS (a relaxed version of PFS) among items in terms of their exposure allocation. \citet{do2021online} consider the problem of auditing the unfair behavior of recommender systems with respect to EF among users. Although these studies formulating versions of EF in the context of recommender systems are related, our contributions are unique in several ways. First, we focus on achieving fairness of \textit{impact among individual items}. In contrast, \citet{patro2020fairrec} consider the item fairness in terms of \textit{exposure} and \citet{do2021online} consider only the \textit{user-side} fairness. In the following sections, we argue that the exposure-based fairness can produce a substantial amount of envy and a highly disparate impact. We are the first to formulate and achieve impact-based fairness rather than fairness-of-exposure in the context of individual fairness in ranking. Second, we build on the exact notions of EF and PFS, while \citet{patro2020fairrec} relax these fairness criteria and rely on EF1 and MMS. We can formulate fairness in rankings as the problem of fairly allocating \textit{divisible} goods by considering a class of \textit{stochastic} ranking policies. Stochastic ranking policies have been adapted in the fair ranking literature~\citep{singh2018fairness,singh2019policy,diaz2020evaluating,yadav2021policy}, and we rely on this stochastic formulation aiming for more desirable fairness criteria than those of previous studies. Finally, our formulation is aware of the position bias in a ranking. Related studies~\citep{patro2020fairrec,do2021online} do not take into account position bias, which is an inevitable factor in dealing with real-world ranking interfaces~\citep{joachims2017unbiased,su2021optimizing}.

\section{An Axiomatic Approach to Individual Fairness in Ranking}
We consider the following class of ranking problems, where a two-sided market platform has responsibilities to both users and items. 
\begin{itemize}
    \item On a news platform, each reader (user) receives a ranking of news stories (items) as a digest of the previous day.
    \item On a hiring platform, each employer (user) receives a weekly ranking of the most relevant job candidates (items) that were added to the database.
    \item For a scientific conference, each reviewer (user) receives a ranking of the most relevant submissions (items) during the bidding process.
\end{itemize}
All problems have in common that the platform not only aims to maximize the utility of the rankings to the users, but that it also needs to be fair to the items. Specifically, the news providers would like a fair share of the traffic for their items, the job candidates deserve to get an adequate number of connections to relevant employers, and every paper should be given an appropriate chance of finding knowledgeable reviewers. 

All these problems are instances of batch-ranking problems, where $\calU=[m]$ is a set of users, and a ranking policy $\pi$ aims to optimally order a set of items $\calI=[n]$ for each user $u \in \calU$. We assume the full-information setting, where we have access to the true relevance labels (or their predictions) $r(u,i) \in \mathbb{R}_{+}$ for all user-item pairs. A ranking $\sigma$ is a permutation of the set of items sampled from a ranking policy $\pi(\cdot|u)$, which is a distribution over all possible rankings of the  items. We consider the general case of stochastic ranking policies, which includes deterministic ranking policies as a special case. Stochastic ranking policies have been utilized in fair ranking research to have fine-grained control over the expected exposure allocation~\citep{singh2019policy,yadav2021policy,diaz2020evaluating}.

\subsection{Utility to Users}

Like in conventional ranking frameworks, we measure the utility that a policy $\pi$ provides to the users through a utility function of the following form.
\begin{align}
    U(\pi) & := \sumU \mE_{\sigma \sim \pi(\cdot|u)} \left[\sumI e(\sigma(i)) \cdot r(u,i) \right]
    \label{eq:user_util}
\end{align}
$\sigma(i)$ is the rank of item $i$ in ranking $\sigma$, and $e(\cdot)$ casts the rank to the exposure probability according to the position-based model (PBM)~\citep{craswell2008experimental}. This definition of utility captures widely-used additive ranking metrics through the choice of $e(\cdot)$. For example, when we define $e(k) := \frac{\mathbb{I}\{k \le K\}}{\log_2(k+1)}$, the utility becomes DCG@K~\citep{jarvelin2002cumulated}. It is also possible to learn an application-specific $e(\cdot)$ from the data~\citep{joachims2017unbiased}. We refer to a ranking policy that maximizes the utility to the users as a utility-maximizing policy, which is defined as $\pimax := \argmax_{\pi \in \Pi} U(\pi)$.

To avoid optimizing over the exponentially large space of rankings, we utilize the fact that only the exposure probability $e(\cdot)$ depends on the rank. This allows us to more concisely express the user utility as follows.
\begin{align}
    U(\pi) := \sumU \sumI \sumK e(k) r(u,i) \allocation \label{eq:user_util_mat}
\end{align}
where $X^{\pi}$ is a \textit{ranking tensor} whose $(u,i,k)$ element $\allocation := \mP(\sigma(i) = k|\pi, u)$ denotes the marginal probability of item $i$ being ranked at the $k$-th position for user $u$ under policy $\pi$,\footnote{$\mP(\sigma(i) = k|\pi, u)= \mE_{\sigma \sim \pi(\cdot|u)} [\mathbb{I} \{\sigma(i) = k\} ] $ where $\mathbb{I}\{\cdot\}$ is the indicator function.} and $X_{u,*,*}^{\pi}$ should be the doubly stochastic matrix\footnote{The sum of each row and column is 1.}. The benefit of using this representation is to reduce the number of parameters needed to specify the effect of policy $\pi$. Specifically, for user $u$, we use only $|\calI|^2$ parameters rather than the exponential number of possible rankings, as all stochastic ranking policies with the same matrix have the same user utility. This will allow us to optimize in the space of doubly stochastic matrices. Given a doubly stochastic matrix, the Birkhoff-von Neumann (BvN) decomposition can be used to efficiently find a stochastic ranking policy that corresponds to the  matrix~\citep{birkhoff1940lattice,singh2018fairness}.

\begin{table*}
\caption{A Illustrative Ranking Problem showing the Exposure Allocations and Impact Fairness of Different Ranking Policies} 
\label{tab:toy_example}
\vspace*{-3mm}
\begin{subtable}{.2\linewidth}
\centering
\caption{True Relevance Table} \label{tab:true_rel_table}
    \scalebox{1.1}{\begin{tabular}{c|cc}
    \toprule
     & $i_1$ & $i_2$ \\ \midrule
    $r(u_1,i)$ & 0.8 & 0.3  \\
    $r(u_2,i)$ & 0.5 & 0.4  \\ \midrule
    $\merit_i$ & 1.3 & 0.7 \\ \bottomrule
  \end{tabular}}
\end{subtable}
\begin{subtable}{.25\linewidth}
\centering
\caption{Max: $U(\pimax)=1.30$} \label{tab:pimax}
    \scalebox{0.95}{\begin{tabular}{c|cc}
    \toprule
      & $i_1$ & $i_2$ \\ \midrule
    $\expo_i(X^{\pimax}|u_1)$ & 1.0 & 0.0  \\
    $\expo_i(X^{\pimax}|u_2)$ & 1.0 & 0.0  \\ \midrule
    $\expo_i(X^{\pimax})$ & 2.0 & 0.0 \\ \midrule
    $\imp_i(X^{\pimax}_{*,i,*})$ & 1.3 & 0.0 \\ \bottomrule
  \end{tabular}}
\end{subtable}
\begin{subtable}{.25\linewidth}
\centering
\caption{Fair: $U(\piexpo)=1.23$} \label{tab:pifair}
    \scalebox{0.95}{\begin{tabular}{c|cc}
    \toprule
      & $i_1$ & $i_2$ \\ \midrule
    $\expo_i(X^{\piexpo}|u_1)$ & 1.0 & 0.0  \\
    $\expo_i(X^{\piexpo}|u_2)$ & 0.3 & 0.7  \\ \midrule
    $\expo_i(X^{\piexpo})$ & 1.3 & 0.7 \\ \midrule
    $\imp_i(X^{\piexpo}_{*,i,*})$ & 0.95 & 0.28 \\ \bottomrule
  \end{tabular}}
\end{subtable}
\begin{subtable}{.25\linewidth}
\centering
\caption{Uniform: $U(\piunif)=1.00$} \label{tab:piunif}
    \scalebox{0.95}{\begin{tabular}{c|cc}
    \toprule
     & $i_1$ & $i_2$ \\ \midrule
    $\expo_i(X^{\piunif}|u_1)$ & 0.5 & 0.5  \\
    $\expo_i(X^{\piunif}|u_2)$ & 0.5 & 0.5  \\ \midrule
    $\expo_i(X^{\piunif})$ & 1.0 & 1.0 \\ \midrule
    $\imp_i(X^{\piunif}_{*,i,*})$ & 0.65 & 0.35 \\ \bottomrule
  \end{tabular}}
\end{subtable}
\end{table*}

\subsection{Impact on Items}

While sorting the items by their probability of relevance maximizes the utility to the users~\cite{robertson1977probability}, many existing works have noted that this naive treatment can lead to rankings that are unfair to the items~\citep{singh2019policy,yadav2021policy,biega2018equity}. In particular, \citet{biega2018equity} measure similarity between individual items by their \textit{amortized merit}, and propose a notion of fairness requiring that \textit{amortized exposure} should be distributed proportional to their amortized merit. This definition of fairness aims at allocating exposure similarly between items with similar merit. However, as we have already argued, this formulation lacks a clear justification for why exposure should be linked proportional to relevance -- or linked via any other specific function. Furthermore, the items only indirectly care about exposure, and they more directly care about the {\em impact} the ranking has on them. For example, a ranking policy that predominantly shows men for high-paying jobs and shows women for low-paying jobs can perfectly obey fairness of exposure, but it clearly violates fairness of impact.

We therefore focus on fairness of impact in this work, where {\em impact} quantifies the effect that a ranking policy has on a specific item $i$. To define impact, we build on an \textit{item-centric} version of the matrix $X^{\pi}_{*,i,*}$ whose $(u,k)$ element is: $\allocation=\mP(\sigma(i) = k|\pi, u)$. This matrix characterizes the \textit{allocation of positions}, i.e., how big a fraction of the $k$-th position in a ranking for user $u$ goes to item $i$. With this notation, we define the \textit{impact} on each item as
\begin{align*}
    \imp_i (X^{\pi}_{*,i,*}) := \sumU \sumK v_i(u,k)  \allocation
\end{align*}
$v_i(u,k)$ is an application-dependent impact function, which defines how much impact (e.g., expected clicks, bookings, revenue) item $i$ receives when it is ranked at the $k$-th position for user $u$. For ease of exposition, we will use the number of relevant users who see each item under the PBM as the impact function, i.e., $v_i(u,k):= e(k)r(u,i)$.
Note that even with this specialization, impact remains more meaningful than fairness of exposure, which does not differentiate between exposure to relevant and non-relevant users. Note that the user utility is equal to the sum of impacts under the specialization
$$U(\pi) = \sumI \imp_i(X^{\pi}_{*,i,*}),$$
making both equally measurable and comparable.

\subsection{Fairness Axioms}

We are now in a position to state axioms that ensure individual fairness with respect to the impact on each item as quantified by $\imp_i(X^{\pi}_{*,i,*})$. Making connections to well-established principles of fair division~\citep{brandt2016handbook,foley1966resource,varian1974equity,kroer2019scalable}, we treat the rank positions as the limited resource to allocate, and impact $\imp_i(X^{\pi}_{*,i,*})$ as the natural valuation of the items for each resource allocation. We start with the axiom of \textit{envy-freeness}.

\begin{definition} \textit{(Envy-Freeness)
A ranking policy $\pi$ is said to be \textit{envy-free} if no item prefers another item's position allocation of $\pi$, i.e., $\imp_i(X^{\pi}_{*,i,*}) \ge \imp_i(X^{\pi}_{*,j,*}), \; \forall i,j \in \calI$.}
\end{definition}

An envy-free ranking policy ensures that no item would prefer the positions allocated to another item over their own allocation. 
If an allocation is envy-free, no item will want to switch allocations with another item. 
Note that envy-freeness can be seen as a natural generalization of fairness-of-exposure as discussed in Appendix~\ref{app:connection}.

The second axiom is \textit{dominance over uniform ranking}, which requires that a fair ranking policy does not provide worse impact $\imp_i(X^{\pi}_{*,i,*})$ to any item than $\piunif$, which samples every possible permutation uniformly at random (uniform ranking policy).

\begin{definition} \textit{(Dominance over Uniform Ranking)
A ranking policy $\pi$ is said to dominate the uniform random policy $\piunif$ if $\pi$ provides better or equal impact on every item compared to $\piunif$, i.e, $\imp_i(X^{\pi}_{*,i,*}) \ge \imp_i(X^{\piunif}_{*,i,*}), \, \forall i \in \calI$. Moreover, at least one item should gain impact strictly better than that under $\piunif$.}
\end{definition}

We regard $\piunif$ as a baseline policy, because it provides an impact distribution that one can achieve without implementing any optimization or policy learning procedure. If a policy makes some items worse off than under $\piunif$, it inflicts active harm on those items and they are likely to abandon the platform.

In addition to the aforementioned fairness desiderata, we also want the ranking policy to be \textit{optimal} in the following sense.

\begin{definition} \textit{(Pareto Optimality)
A ranking policy $\pi$ is said to be \textit{Pareto optimal} if no alternative policy can make some items strictly better off without making any others strictly worse off.}
\end{definition}

Pareto optimality ensures that the user utility, and equivalently the sum of item impacts, cannot easily be improved. This codifies that the policy should not needlessly sacrifice utility to the users or aggregate impact to the items, and that there is no obviously avoidable harm to any of the items.

\subsection{Fairness of Exposure Violates Axioms}
To motivate the need for new algorithms to achieve fairness of impact, we now show in detail that the conventional fairness-of-exposure framework can cause envy among the items and create an unfair impact distribution in light of our axioms.
Exposure-based fairness imposes a constraint to allocate exposure proportional to the amortized merit~\citep{biega2018equity}. Here, we describe a more general constraint~\citep{wang2021fairness}, where merit is
quantified through an application-dependent link function $f(\cdot)>0$,
\begin{align}
    \frac{\expo_i (X^{\pi})}{f(\merit_i)} = \frac{\expo_j (X^{\pi})}{f(\merit_j)} , \quad \forall i,j \in \calI,
    \label{eq:exposure_based_constraint}
\end{align}
where $\merit_i := \sumU  r (u,i)$ is the \textit{amortized merit}, i.e., the relevance of item $i$ amortized over all users. In contrast, $\expo_i (X^{\pi}) := \sum_u \expo_i(X^{\pi}|u)  = \sum_u \sum_k e(k) \allocation$ is the total amount of exposure allocated to item $i$ under policy $\pi$, i.e., the \textit{amortized exposure}. The link function $f(\cdot)$ maps the amortized merit to a positive merit value. A typical choice is $f(x)=x$, which leads to a linear program to optimize the rankings~\citep{singh2018fairness,singh2019policy,biega2018equity}.

Algorithms for computing exposure-based fair rankings find a doubly stochastic matrix that maximizes the utility to the users (or equivalently the sum of impacts), subject to Eq.~\eqref{eq:exposure_based_constraint} as follows.
\begin{align*}
    \piexpo \quad & \\
    = \argmaxI & \sumU \sumI \sumK e(k) r(u,i) \allocation \; \, \left(= \sumI \imp_i(X_{*,i,*}^{\pi}) \right), 
    \\
     \mathrm{s.t.} \quad & \frac{\expo_i(X^{\pi})}{f(\merit_i)} = \frac{\expo_j(X^{\pi})}{f(\merit_j)} , \quad \forall i,j \in \calI, \\
      & \; \sumK \allocation = 1, \; \forall (u, i) \\
      & \; \sumI \allocation = 1, \; \forall (u, k)
       \\ \quad & 0 \le \allocation \le 1,\; \forall (u,i,k) 
\end{align*}
We use $\piexpo$ to denote a policy that solves the above optimization and call it the exposure-based fair ranking policy.

While widely used as the desideratum for fair ranking~\citep{singh2018fairness,biega2018equity,singh2019policy,yadav2021policy,zehlike2020reducing,morik2020controlling,yang2021maximizing,jeunen2021top,diaz2020evaluating}, the following provides a counterexample which illustrates that a fair allocation of exposure cannot avoid a \textit{disparate impact} among individual items.
Our counterexample consists of two users ($\calU=\{u_1,u_2\}$) and two items ($\calI=\{i_1,i_2\}$). We know the true relevance labels of every user-item pair, which is given in Table~\ref{tab:true_rel_table}. For simplicity, we also assume that only the top-ranked item is exposed to the users ($e(1)=1, e(2)=0$). We consider three ranking policies, \textbf{(i)} the utility-maximizing policy $\pimax$, \textbf{(ii)} the exposure-based fair ranking policy $\piexpo$, and, \textbf{(iii)} the uniform random policy $\piunif$.

Tables~\ref{tab:pimax} to~\ref{tab:piunif} provide the exposure allocation of the three policies and how much impact each item receives. $\pimax$ allocates all exposure on the item having highest relevance ($i_1$). As a result, it achieves the highest user utility ($U(\pimax)=1.3$) while leading to the unfair situation where $i_2$ obtains no exposure despite its substantial relevance.

$\piexpo$ deals with this unfair allocation by imposing the exposure constraint. Specifically, it allocates some amount of exposure to the less relevant item ($i_2$) and ensures that the exposure is allocated proportional to merit as shown in Table~\ref{tab:pifair}.
However, our example indicates that $\piexpo$ violates  impact-based fairness. First, $\piexpo$ is \textit{not envy-free}. Indeed, $i_2$ envies the allocation of $i_1$, because impact on $i_2$ can be improved by swapping its allocation with that of $i_1$.\footnote{We know that $\imp_{i_2}(X^{\piexpo}_{*,\red{i_2},*})=0.28$ from Table~\ref{tab:pifair}. If we replaced $X^{\piexpo}_{*,\red{i_2},*}$ with $X^{\piexpo}_{*,\blue{i_1},*}$, impact on $i_2$ would increase, i.e., $\imp_{i_2}(X^{\piexpo}_{*,\blue{i_1},*})=0.42$.} Second, $\piexpo$ does \textit{not dominate} $\piunif$ for all items. It is easy to see that impact on $i_2$ under $\piexpo$ (0.28) is smaller than under $\piunif$ (0.35). This implies that $\piexpo$ improves impact on $i_1$ at the cost of impact on $i_2$, creating a substantial disparity. 

Our counterexample indicates that exposure-based fairness allows a policy to produce envy and an unfair distribution of impact. The following theorem more formally illustrates the possible disparate impact of the existing exposure-based framework.

\begin{theorem} \label{thm:sub_optimality}
There exist two-sided markets with $|\calU|=|\calI|=n$ such that, even under the exposure-based fair ranking policy $\piexpo$ with $f(x)=x$, there exists item $i$ whose maximum envy grows while impact compared to $\piunif$ diminishes with the market size $n$, i.e., $$\frac{\max_{j \in \calI} \imp_i(X^{\piexpo}_{*,j,*})}{\imp_i(X^{\piexpo}_{*,i,*})} = \Omega(n), \; \frac{\imp_i(X^{\piexpo}_{*,i,*})}{\imp_i(X^{\piunif}_{*,i,*})} = \mathcal{O}(n^{-1}).$$
for any $e(\cdot)$ such that the exposure-fair constraint in Eq.~\eqref{eq:exposure_based_constraint} is feasible.
\end{theorem}

Theorem~\ref{thm:sub_optimality} indicates that exposure-based fair rankings can produce large amounts of envy for some items. It can also actively harm some items beyond a reasonable baseline. \citet{singh2018fairness} describe the \textit{disparate impact} constraint aiming at allocating impact proportional to merit in the context of \textit{group} fairness. However, this constraint merely leads to the uniform ranking in the case of \textit{individual} fairness~\citep{singh2018fairness}.
The disparate impact caused by the exposure-based framework and the inapplicability of the disparate impact constraint validate that our framework of \textit{impact-based} individual fairness is fundamentally new and different from conventional exposure-based fairness.

\section{Computing Rankings with Fairness of Impact Guarantees}

While the previous section introduced the axiomatization of fairness-of-impact for ranking and showed that the conventional fairness-of-exposure approach does not satisfy these axioms, it is not yet clear whether an efficient algorithm exists to compute fair rankings -- and whether fair rankings even exist. To approach this algorithmic question, we formulate the following optimization problem which extends the concept of \textit{Nash Social Welfare} (NSW) to fair rankings. 
\begin{align*}
    \pinsw = \argmaxI \quad & \prod_{i \in \calI} \imp_i(X_{*,i,*}^{\pi}), \\
     \mathrm{s.t.} \quad & 
      \sumK \allocation = 1, \; \forall (u, i) \\
      & \sumI \allocation = 1, \; \forall (u, k)
       \\ \quad & 0 \le \allocation \le 1,\; \forall (u,i,k) 
\end{align*}
The objective maximizes the \textit{product} of impacts on the items, and the constraints express that each item needs to have probability 1 to be placed in some position, and each position needs to have probability 1 of receiving an item. This constraint structure makes this optimization problem different from the standard NSW in fair division~\citep{eisenberg1959consensus,birnbaum2011distributed,kroer2019scalable}. However, we retain that the optimization problem is convex and thus efficiently solvable, which is easy to see by equivalently replacing the product of impacts in objective with the sum of their logarithms, $\sumI \log \imp_i(X_{*,i,*}^{\pi})$.\footnote{This type of program is know as the Eisenberg-Gale (EG) program~\citep{eisenberg1959consensus}.} We call the policy that solves the optimization problem the NSW policy, which we denote by $\pinsw$.

A useful intuition is that the NSW does not allow any item to have zero impact, thus achieving a more equitable impact distribution. In particular, the NSW becomes zero if there is even a single item with zero impact. This contrasts with the conventional objective of maximizing user utility (which equals the sum of impacts), which may be maximized even if there are some items receiving no impact. The following shows more formally that this intuition is correct, and that $\pinsw$ guarantees (approximate) dominance over $\piunif$. Furthermore, the following shows that $\pinsw$ is (approximately) envy-free and Pareto optimal.

\begin{theorem} \label{thm:nsw}
If $K=1$ and only the top-ranked item contributes to the utility and impact (e.g., DCG@1), $\pinsw$ is exactly Pareto optimal, is envy-free, and dominates $\piunif$. For the more general case with $K>1$, $\pinsw$ is still Pareto optimal. Moreover, let us call a pair of items $i$ and $j$ ``$\epsilon$-twins'' if $\max_{u}|r(u,i)-r(u,j)| \le \epsilon$. Then, if every item has at least $K+1$ $\epsilon$-twins and $n/m = \calO(1)$, we have
\begin{align*}
    \max_{j \in \calI} \, \imp_i(X^{\pinsw}_{*,j,*}) - \imp_i(X^{\pinsw}_{*,i,*}) = \calO(\epsilon), \notag \\
     \imp_i(X^{\piunif}_{*,i,*})  - \imp_i(X^{\pinsw}_{*,i,*})  = \calO(\epsilon),
\end{align*}
for all item $i \in \calI$.
\end{theorem}

In the case of $K=1$, the solution of our problem coincides with the allocation computed via the \textit{competitive equilibrium from equal incomes (CEEI)}, a classic model of market equilibrium~\citep{varian1974equity,budish2011combinatorial}, which is known to be Pareto optimal, to be envy-free, and to dominate a uniform random allocation. When $K>1$, the classical result from CEEI does not necessarily hold due to the additional constraints in our problem~\citep{kroer2019scalable}. However, Theorem~\ref{thm:nsw} characterizes that, even for the general case, $\pinsw$ can approximately satisfy our fairness axioms if there are items that are similar to each other, which is reasonable for large markets.\footnote{Note that it may be reasonable to assume that $\epsilon$ decreases with the size of the market, since it should be easier to find a sufficient number of $\epsilon$-twins in a larger market. This means we could expect $\pinsw$ to more accurately satisfy our axioms with an increasing number of items.} Note that Theorem~\ref{thm:nsw} assumes the availability of the true relevance labels, but in Section~\ref{sec:experiment}, we empirically show that $\pinsw$ robustly produces fair distributions of impact even with predicted relevance values.

Table~\ref{tab:pinsw} illustrates these guarantees of $\pinsw$ on the same example used earlier. First, we can see that $\pinsw$ produces an \textit{envy-free} allocation of positions, i.e., neither item prefers the other item's allocation. In addition, $\pinsw$ dominates $\piunif$, since every item gains higher impact than under $\piunif$.

\begin{table}[t]
\caption{Allocation of the NSW Policy: $U(\pinsw)=1.20$} \label{tab:pinsw}
\vspace{-2mm}
\centering
\scalebox{0.95}{\begin{tabular}{c|cc}
\toprule
  & $i_1$ & $i_2$ \\ \midrule
$\expo_i(X^{\pinsw}|u_1)$ & 1.0 & 0.0  \\
$\expo_i(X^{\pinsw}|u_2)$ & 0.0 & 1.0  \\ \midrule
$\expo_i(X^{\pinsw})$ & 1.0 & 1.0 \\ \midrule
$\imp_i(X^{\pinsw}_{*,i,*})$ & 0.8 & 0.4 \\ \bottomrule
\end{tabular}}
\end{table}

\begin{table}[t]
\caption{Exposure Allocations and Impact Fairness of the $\alpha$-NSW Policies with Different Hyper-parameter Values} \label{tab:pinsw_alpha}
\vspace{-3mm}
\begin{subtable}{.49\linewidth}
\centering
\caption{1-NSW: $U(\pi_{\mathrm{1\textit{-NSW}}})=1.21$} \label{tab:pinsw_1}
    \scalebox{0.95}{\begin{tabular}{c|cc}
    \toprule
      & $i_1$ & $i_2$ \\ \midrule
    $\expo_i(X^{\pi_{\mathrm{1\textit{-NSW}}}}|u_1)$ & 1.0 & 0.0  \\
    $\expo_i(X^{\pi_{\mathrm{1\textit{-NSW}}}}|u_2)$ & 0.1 & 0.9  \\ \midrule
    $\expo_i(X^{\pi_{\mathrm{1\textit{-NSW}}}})$ & 1.1 & 0.9 \\ \midrule
    $\imp_i(X^{\pi_{\mathrm{1\textit{-NSW}}}}_{*,i,*})$ & 0.85 & 0.36 \\ \bottomrule
  \end{tabular}}
\end{subtable} 
\vspace{-3mm}
\begin{subtable}{0.49\linewidth}
\centering
\caption{2-NSW: $U(\pi_{\mathrm{2\textit{-NSW}}})=1.24$} \label{tab:pinsw_2}
    \scalebox{0.95}{\begin{tabular}{c|cc}
    \toprule
     & $i_1$ & $i_2$ \\ \midrule
    $\expo_i(X^{\pi_{\mathrm{2\textit{-NSW}}}}|u_1)$ & 1.0 & 0.0  \\
    $\expo_i(X^{\pi_{\mathrm{2\textit{-NSW}}}}|u_2)$ & 0.4 & 0.6  \\ \midrule
    $\expo_i(X^{\pi_{\mathrm{2\textit{-NSW}}}})$ & 1.4 & 0.6 \\ \midrule
    $\imp_i(X^{\pi_{\mathrm{2\textit{-NSW}}}}_{*,i,*})$ & 1.00 & 0.24 \\ \bottomrule
  \end{tabular}}
\end{subtable}
\end{table}

\subsection{Controlling the Fairness/Utility Trade-Off}

In some situations, we may want to exert explicit control over how much user utility to sacrifice for stronger fairness guarantees to the items.
For example, in Tables~\ref{tab:pinsw} and~\ref{tab:toy_example}, the user utility of $\pinsw$ (1.20) is lower than that of $\pimax$ (1.30) and $\piexpo$ (1.23), and we may want to put more emphasis on user utility. To enable explicit control, we extend the NSW to what we call the $\alpha$-NSW as follows.
\begin{align*}
    \pinswalpha = \argmaxI \quad & \prod_{i \in \calI} \imp_i(X_{*,i,*}^{\pi})^{\merit_i^{\,\red{\alpha}}}, \\
     \mathrm{s.t.} \quad  
      & \sumK \allocation = 1, \; \forall (u, i) \\
      & \sumI \allocation = 1, \; \forall (u, k)
      \\ \quad & 0 \le \allocation \le 1,\; \forall (u,i,k) 
\end{align*}
$\alpha \ge 0$ is a hyper-parameter that controls the balance between maximizing user utility (large $\alpha$) and guaranteeing impact-based fairness (small $\alpha$). 
When $\alpha=0$, the $\alpha$-NSW objective is reduced to the standard NSW. Note that we can again take the logarithm to rewrite the objective to $\sumI \merit_i^{\,\red{\alpha}} \log \imp_i(X_{*,i,*}^{\pi})$ and solve the program efficiently. The $\alpha$-NSW policy ensures only a weaker version of envy-freeness and dominance over $\piunif$ as detailed in the appendix, which means it can achieve higher user utility compared to the standard NSW.

Table~\ref{tab:pinsw_alpha} illustrates the allocations of the $\alpha$-NSW policy $\pinswalpha$ for $\alpha=1$ and $\alpha=2$ on the example problem. We see that a larger value of $\alpha$ leads to a better user utility, i.e., $U(\pi_{\textit{NSW}}) < U(\pinswone) < U(\piexpo) < U(\pi_{2\textit{-NSW}})$, while a smaller value leads to a more balanced impact distribution. What is particularly notable here is that $\pinswone$ still dominates $\piunif$ and remains envy-free, while achieving a better utility compared to $\pinsw$.
This suggests that an appropriate choice of $\alpha$ may allow us to improve user utility over $\pinsw$ without sacrificing envy-freeness and dominance over $\piunif$ on particular problem instances, even if we no longer have a-priori guarantees that hold over all problem instances. In the next section, we empirically explore how $\alpha$ controls this trade-off.

\section{Experiments} \label{sec:experiment}
We first present experiments on synthetic data where we can control the popularity pattern in the market and the accuracy of the relevance prediction. In addition, we use real-world extreme classification datasets to evaluate how our method works with realistic relevance predictions. Our experiment implementation is available at \href{https://github.com/usaito/kdd2022-fair-ranking-nsw}{\blue{https://github.com/usaito/kdd2022-fair-ranking-nsw}}.

\begin{figure*}[ht]
\includegraphics[width=17.5cm]{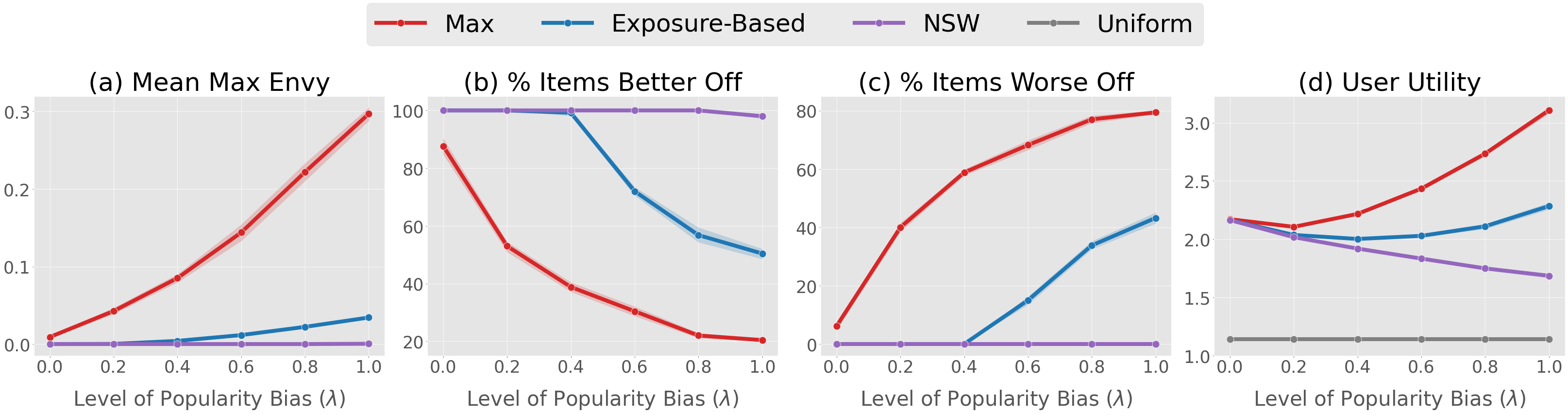}
\vspace*{-2mm}
\caption{Fairness and user utility of the NSW policy compared to baseline policies on synthetic data for varying levels of popularity bias.}
\label{fig:pop}
\end{figure*}

\begin{figure*}[ht]
\includegraphics[width=17.5cm]{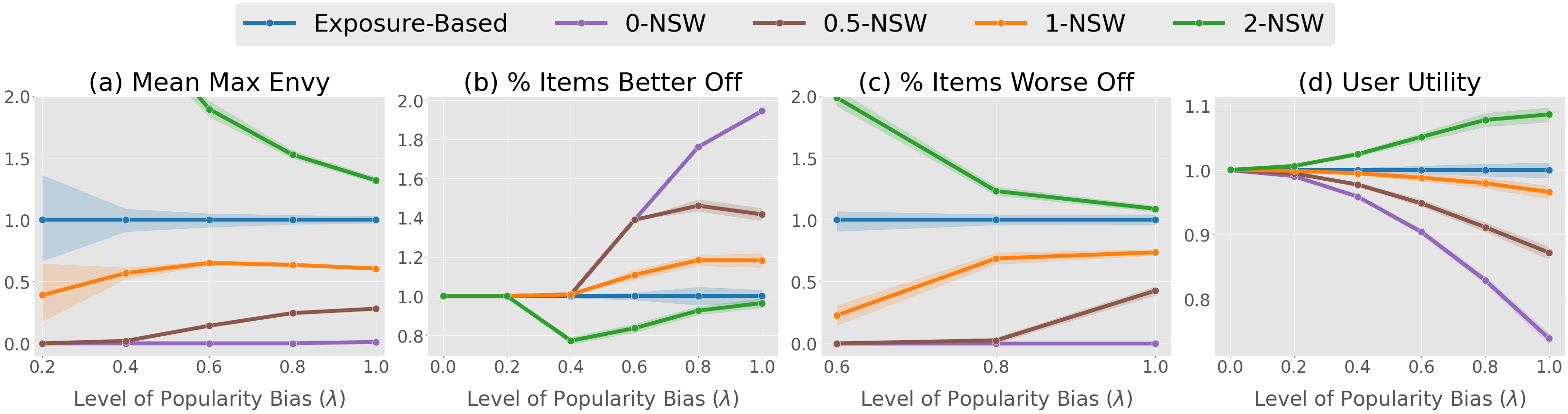}
\vspace*{-2mm}
\caption{Fairness and user utility relative to the exposure-based fair policy $\piexpo$ for various $\alpha$-NSW policies. Note that we omit $\lambda=0.0$ from (a) and $\lambda=0.0,0.2,0.4$ from (c), because the measures of $\piexpo$ are (almost) zero for these cases.}
\label{fig:alpha}
\end{figure*}

\subsection{Synthetic Data} \label{sec:synthetic}
To generate synthetic data, we first define the ground-truth relevance between user $u$ and item $i$ as
\begin{align}
    \reltrue(u,i) = (1 - \lambda) \cdot r_{\textit{unif}}\,(u,i) + \lambda \cdot r_{\textit{pop}}(u,i),
    \label{eq:pop}
\end{align}
where $r_{\textit{unif}}\,(u,i)$ is an independent and uniform draw within range $[0,1]$. The second term infuses \textit{popularity bias}, which we define as
\begin{align*}
    r_{\textit{pop}}(u,i) 
    & := \begin{cases}
        \frac{n-i+1}{n} \cdot \frac{m-u+1}{m} & (\textit{randomly sampled 70\% of the items})\\
        \frac{n-i+1}{n} \cdot \frac{u}{m} & (\textit{rest of the items})
    \end{cases}
\end{align*}
$\lambda \in [0,1]$ controls the popularity pattern, where a larger value of $\lambda$ leads to more severe popularity bias. We also simulate that we typically only have access to (inaccurately) predicted relevances, which we model as
\begin{align*}
    \relpred (u,i) := \mathrm{clip} \big(\reltrue (u,i) + \eta, 0, 1 \big), \quad \eta \sim \mathrm{Unif} (-c,c),
\end{align*}
where $\eta$ is independent and uniform noise, and $c \ge 0$ controls the accuracy of the predicted relevance values. We use the predicted relevance $\relpred (u,i)$ to optimize the rankings, and evaluate the item fairness and user utility of the policies using the ground-truth relevance $\reltrue(u,i)$. We use the inverse examination function $e(k) := \mathbb{I}\{k \le K\}/k$ in the definition of utility and impact, but we also present experiments with a different function in the appendix. 

We evaluate the degree of individual item fairness of a ranking policy $\pi$ using the following measures:
\begin{enumerate}
    \item Mean Max Envy (smaller is better): $$\frac{1}{|\calI|} \left\{ \sumI \max_{j\in \calI} \, \imp_i(X_{*,j,*}^{\pi}) - \imp_i(X_{*,i,*}^{\pi}) \right\}$$
    \item Proportion (\%) of items for which $\pi$ improves impact by over 10\% compared to $\piunif\,$ (larger is better):   
    $$\frac{100}{|\calI|} \sumI \mathbb{I} \left\{ \imp_i(X_{*,i,*}^{\pi})/\imp_i(X_{*,i,*}^{\piunif}) \ge 1.1 \right\} $$
    \item Proportion (\%) of items for which $\pi$ decreases impact by over 10\% compared to $\piunif\,$ (smaller is better): 
    $$\frac{100}{|\calI|} \sumI \mathbb{I} \left\{ \imp_i(X_{*,i,*}^{\pi})/\imp_i(X_{*,i,*}^{\piunif}) \le 0.9 \right\} $$
\end{enumerate}

The following evaluates $\pinsw$ in comparison to $\pimax$, $\piexpo$, and $\piunif$. We also evaluate how the hyper-parameter $\alpha$ of $\pinswalpha$ controls the trade-off of item fairness and user utility. Note that we use $|\calU|=100$, $|\calI|=50$, $c=0.05$, $\lambda=0.5$, $K=5$ as defaults, and vary each experiment configuration to see how each affects the behavior of the ranking policies.\footnote{The results with varying number of items ($n$) and lengths of ranking ($K$) are reported in Appendix~\ref{app:experiment_detail}} Results are averaged over 10 simulation runs performed with different random seeds.

\begin{figure*}[ht]
\includegraphics[width=17.5cm]{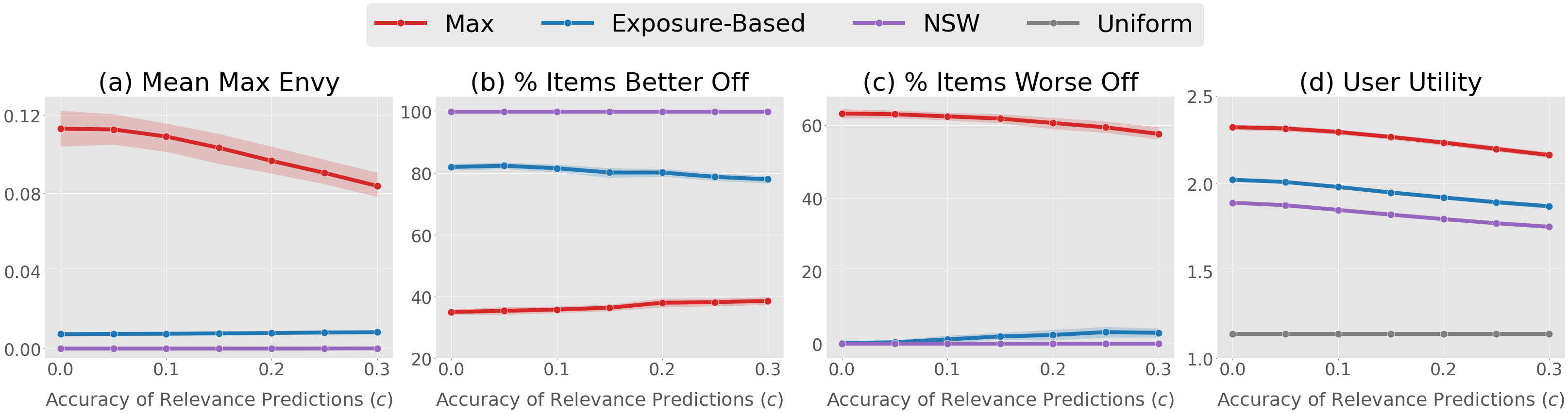}
\vspace*{-2mm}
\caption{Fairness and user utility of the NSW policy compared to baseline policies on synthetic data for varying levels of relevance-prediction accuracy.}
\label{fig:noise}
\end{figure*}

\begin{figure*}[ht]
\includegraphics[width=17.5cm]{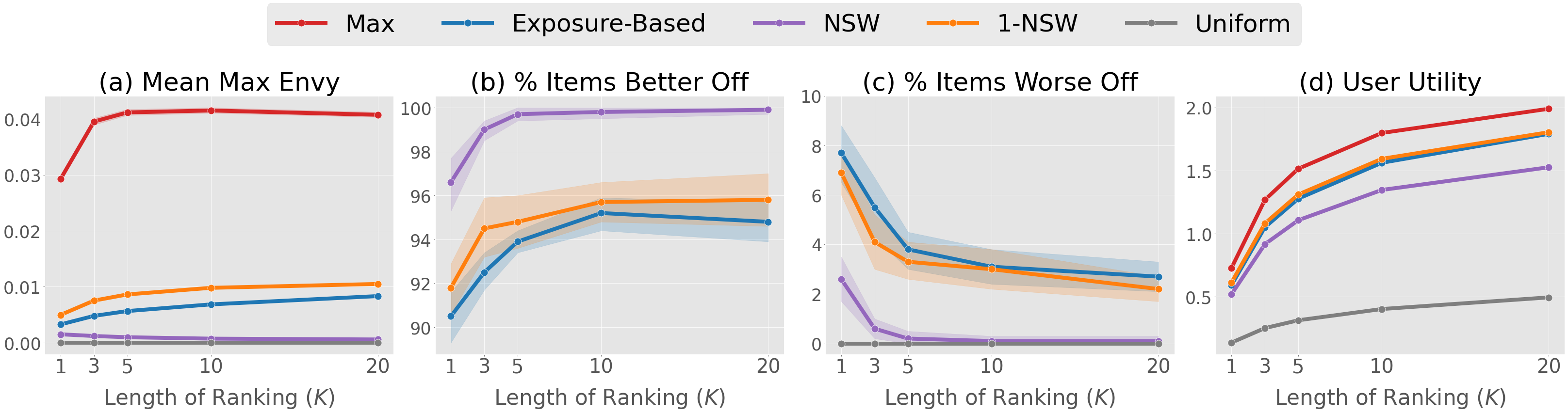}
\vspace*{-2mm}
\caption{Fairness and user utility on the Delicious dataset for varying lengths of the ranking.}
\label{fig:deli}
\end{figure*}

\subsubsection{How do the ranking policies perform with different popularity patterns?}
Here, we test how the fairness measures and user utility of the ranking policies change with different popularity patterns. For this purpose, we vary the experiment parameter $\lambda$ in Eq.~\eqref{eq:pop} in the range of $\{0.0,0.2,\ldots,1.0\}$. First, Figure~\ref{fig:pop} (a) shows the amount of envy produced under different levels of popularity bias. The figure suggests that $\pinsw$ is always almost envy-free, while $\pimax$ and $\piexpo$ produce larger amounts of envy for markets with greater popularity bias. Next, Figure~\ref{fig:pop} (b) shows how many items gain substantially better impact compared to $\piunif$ under different levels of popularity bias. We observe that $\pinsw$ substantially improves impact on almost all items in all cases, while $\pimax$ and $\piexpo$ fail to achieve this desideratum. In addition, Figure~\ref{fig:pop} (c) indicates that $\pimax$ and $\piexpo$ decrease impacts on some items over 10\% from those under $\piunif$. In particular, when $\lambda=1.0$, 80\% and 40\% of the items experience substantial loss of impact under $\pimax$ and $\piexpo$, respectively. Instead, $\pinsw$ does not substantially decrease any item's impact below $\piunif$, even if there exists severe popularity bias. These results demonstrate that $\pinsw$ avoids producing envy and provides fair improvements in impact for almost all items. On the other hand, both $\pimax$ and $\piexpo$ can lead to an unfair impact distribution and produce substantial amounts of envy, especially in the presence of severe popularity bias.

Although $\pinsw$ is the most desirable in terms of item fairness, Figure~\ref{fig:pop} (d) suggests the expected trade-off between maximizing the user utility and satisfying our impact-based axioms. Figure~\ref{fig:pop} (d) shows how much user utility each policy gains under different popularity patterns. We observe that $\pimax$ achieves the largest user utility followed by $\piexpo$ and $\pinsw$. This trend does not change across different levels of popularity bias, however, the gap in user utility among those policies varies. When $\lambda=0$ and there is no popularity bias in the market, all policies except for $\piunif$ achieve almost the same user utility. However, for larger values of $\lambda$, the gap in user utility between $\pimax$, $\piexpo$, and $\pinsw$ becomes greater. In particular, $\pimax$ and $\piexpo$ improve user utility, while that of $\pinsw$ decreases. An explanation is that an increase in popularity bias makes it easier to maximize the user utility by allocating large amounts of exposure to popular items. Conversely, $\pinsw$ shows the opposite trend since it needs to ever more strongly counteract the popularity bias to maintain impact fairness.

\subsubsection{How does the hyper-parameter of $\pinswalpha$ balance item fairness and user utility?}
This experiment investigates how effectively one can control the balance between item fairness and user utility through the hyper-parameter $\alpha$ of $\pinswalpha$. To this end, we evaluate $\pinswalpha$ with $\alpha \in \{0.0, 0.5, 1.0, 2.0\}$ (when $\alpha=0$, $\pinswalpha$ is identical to $\pinsw$) in comparison to $\piexpo$. Figure~\ref{fig:alpha} reports the fairness measures and user utility of $\pinswalpha$ relative to those of $\piexpo$. Overall, the result demonstrates that $\pinswalpha$ is able to choose a range of trade-offs through the hyper-parameter. The most interesting trade-off is achieved with $\alpha=1.0$. Figure~\ref{fig:alpha} (a) indicates that $\pinswone$ reduces the amount of envy over 40\% compared to $\piexpo$ in all cases. Moreover, Figure~\ref{fig:alpha} (b) and (c) demonstrate that $\pinswone$ provides a far more equitable distribution of impact compared to $\piexpo$. Finally, Figure~\ref{fig:alpha} (d) shows that $\pinswone$ achieves almost the same user utility as $\piexpo$ (about 4\% drop even when $\lambda=1.0$). This result suggests that $\pinswalpha$ has the potential to achieve a substantially fairer impact distribution than $\piexpo$ while achieving a comparable level of user utility.

\subsubsection{How do the ranking policies perform under different relevance-prediction accuracies?}
We also empirically investigate how robust the ranking policies are to inaccurate relevance labels, given that one would typically use predicted relevances in real-world applications. Figure~\ref{fig:noise} illustrates fairness and user utility for varying values of the accuracy parameter $c$. The plots show that user utility slightly decreases as the relevance labels become more inaccurate, while fairness remain almost constant for all policies. Notably, $\pinsw$ is almost envy-free and achieves a fairer impact distribution compared to the other policies in all situations, indicating that the NSW approach is robust to inaccurate relevance predictions.

\subsection{Real-World Data}
To further evaluate how our method performs with predicted relevances on real-world data, we adapt two multilabel extreme classification datasets, namely Delicious and Wiki10-31K from the Extreme Classification Repository.\footnote{http://manikvarma.org/downloads/XC/XMLRepository.html} We regard each data as a user and each label as an item. If a data belongs to a label, then they are considered relevant. As preprocessing, we randomly sample 100 labels,\footnote{We sample label $i$ with probability $ \mu \cdot \mathbb{I} \{i \text{ is one of top 100 frequent labels}\} + (1 - \mu) \cdot (100 / |\calI|)$ (where we set $\mu=0.5$) to introduce some popularity bias in the data.} and then split the data into 90\% training and 10\% test sets. We train a Logistic Regression model on the training set, and predict the probabilities of each data belonging to the labels, which correspond to the relevance predictions. Based on the predictions, the ranking policies optimize the ordering of the labels for each data. We finally evaluate item fairness and user utility of the ranking policies on the test set using the true class labels as the ground-truth relevance. The fairness measures are the same as in the previous section. We use the inverse examination function and vary $K \in \{1,3,5,10,20\}$. We iterate the simulation 10 times with different seeds and report the aggregated results.

Figure~\ref{fig:deli} compares fairness and user utility of $\pimax$, $\pinsw$, $\pinswone$, $\piexpo$ and $\piunif$ on the Delicious dataset. First, $\pinsw$ achieves almost envy-free and fair impact distribution with increasing $K$, suggesting that our desiderata can be achieved even with realistically inaccurate relevance predictions. In addition, $\pinswone$ succeeds in finding a utility-fairness trade-off that is more desirable than that of $\piexpo$. Specifically, $\pinswone$ obtains user utility slightly better than $\piexpo$ while leading to a fairer impact distribution. Figure~\ref{fig:deli} (a) suggests that $\pinswone$ produces a larger amount of envy compared to $\piexpo$, but the amount is controllable by tuning $\alpha$. Appendix~\ref{app:experiment_detail} reports qualitatively similar results on Wiki10-31K.

\section{Conclusion}
We provide a new conceptualization of fairness in ranking as a resource allocation problem and propose impact-based axioms for individual item fairness. This allows us to build upon well-established principles of fairness from economics, removing the need to choose a difficult-to-justify link function as required for exposure-based fairness. We also contribute an efficient algorithm for computing impact-fair rankings, adapting the Nash Social Welfare to ranking problems. Furthermore, we develop a practical extension of the NSW, which enables us to control the trade-off between user utility and item fairness. This work opens a wide range of new research directions, ranging from the development of end-to-end policy learning methods for impact-fair ranking, to the use of partial-information data and extensions to sequential ranking problems.

\begin{acks}
This research was supported in part by NSF Awards IIS-1901168 and IIS-2008139. Yuta Saito was supported by the Funai Overseas Scholarship. All content represents the opinion of the authors, which is not necessarily shared or endorsed by their respective employers and/or sponsors. 
\end{acks}

\bibliographystyle{ACM-Reference-Format}
\bibliography{main.bbl}

\appendix

\section{Connecting Exposure-based and Impact-based Notions of Fairness} \label{app:connection}

The existing exposure-based fairness and our impact-based axioms have an interesting connection. Here, we show that the equality of attention of \citet{biega2018equity} is a special case of our notion of envy-freeness.

\begin{theorem} \label{thm:equality_of_attention}
If the impact function is defined as $v_i(u,k):=e(k)$, then envy-freeness coincides with the equality of attention.
\begin{proof}
Consider a pair of items $i,j \in \calI (i \neq j)$. Given that $v_i(u,k):=e(k)$, envy-freeness mandates that
\begin{align}
    \imp_i(X_{*,i,*}^{\pi}) \ge  \imp_i(X_{*,j,*}^{\pi})
    & \Longrightarrow  \expo_i ( X^{\pi}) \ge \expo_j (X^{\pi}). \label{eq:a_1}
\end{align}
Analogously, for the opposite direction, we should have that
\begin{align}
    \imp_j(X_{*,j,*}^{\pi}) \ge  \imp_j(X_{*,i,*}^{\pi}) \Longrightarrow \expo_j (X^{\pi}) \ge \expo_i ( X^{\pi}), \label{eq:a_2}
\end{align}
Thus, from Eqs~\eqref{eq:a_1} and~\eqref{eq:a_2}, envy-freeness coincides with the equality of attention, which requires that the amortized exposure should be equal for all items, i.e., $\expo_i ( X^{\pi}) = \expo_j (X^{\pi}), \; \forall i,j \in \calI$.
\end{proof}
\end{theorem}

Note that $\pinswalpha$ satisfies the following relaxed version of envy-freeness and dominance over $\piunif$ in the case of $K=1$.
\begin{align*}
    \textit{Weighted Envy-Freeness: } \frac{\imp_i(X^{\pi}_{*,i,*})}{\merit_i^{\,\alpha}} \ge \frac{\imp_i(X^{\pi}_{*,j,*})}{\merit_j^{\,\alpha}}, \; \forall j \in \calI, \notag \\
    \textit{Weighted Dominance: } \imp_i(X^{\pi}_{*,i,*}) \ge \frac{n\merit_i^{\,\alpha}}{\sum_j \merit_j^{\,\alpha}} \imp_i(X^{\piunif}_{*,i,*}),
\end{align*}
for all $i \in \calI$. 
Here, we can follow a similar step as in Theorem~\ref{thm:equality_of_attention} and show that weighted envy-freeness coincides with the equity (not equality) of attention~\citep{biega2018equity} when $\alpha=1$ and $v_i(u,k):=e(k)$.

\begin{theorem} \label{thm:equity_of_attention}
If the impact function is defined as $v_i(u,k):=e(k)$, then weighted envy-freeness with $\alpha=1$ coincides with the equity of attention.
\begin{proof}
Consider a pair of items $i,j \in \calI (i \neq j)$. Given that $\alpha=1$ and $v_i(u,k):=e(k)$, weighted envy-freeness mandates that
\begin{align}
    \frac{\imp_i(X_{*,i,*}^{\pi})}{\merit_i} \ge \frac{\imp_i(X_{*,j,*}^{\pi})}{\merit_j}
    & \Longrightarrow  \frac{\expo_i ( X^{\pi})}{\merit_i} \ge \frac{\expo_j (X^{\pi})}{\merit_j}. \label{eq:b_1}
\end{align}
Analogously, for the opposite direction, we should have that
\begin{align}
    \frac{\imp_j(X_{*,j,*}^{\pi})}{\merit_j} \ge \frac{\imp_j(X_{*,i,*}^{\pi})}{\merit_i} \Longrightarrow \frac{\expo_j (X^{\pi})}{\merit_j} \ge \frac{\expo_i ( X^{\pi})}{\merit_i}, \label{eq:b_2}
\end{align}
Thus, from Eqs~\eqref{eq:b_1} and~\eqref{eq:b_2}, weighted envy-freeness coincides with the equity of attention, which requires that the amortized exposure should be allocated proportional to merit for all items (Eq.~\eqref{eq:exposure_based_constraint} in the main text with $f(x)=x$), i.e., $\frac{\expo_i ( X^{\pi})}{\merit_i} = \frac{\expo_j (X^{\pi})}{\merit_j}, \; \forall i,j \in \calI$.
\end{proof}
\end{theorem}

\section{Proofs}

\paragraph{Proof of Theorem~\ref{thm:sub_optimality}}
Given $r(u,i) = (n-u+1)(n-i+1)/n^2$, amortized merit of $i$ is $\merit_i = \sum_u r(u,i) = (n+1)(n-i+1)/2n$.
To satisfy the exposure fairness constraint in Eq.~\eqref{eq:exposure_based_constraint}, exposure-fair ranking policy $\piexpo$ needs to allocate the following amount of exposure to item $i$
\begin{align*}
    \expo_i (X^{\piexpo}) = \frac{\merit_i}{\sum_i \merit_i} n \te = \frac{2(n-i+1)}{n+1} \te
\end{align*}
where $\te :=\sum_k e(k)$ is the total amount of exposure available for each user and we use $\sum_i \merit_i = (n+1)^2/4$.

Then, we have
\begin{align}
    \expo_n (X^{\piexpo}) &= \frac{2\te}{n+1} \notag \\
    \imp_n  (X_{*,n,*}^{\piexpo}) & \le r(1,n) \expo_n (X^{\piexpo}) = \frac{2\te}{n(n+1)} \label{eq:a_3}
\end{align}
Note that Eq.~\eqref{eq:a_3} holds for any policy (including $\piexpo$) that satisfies the exposure-fair constraint in Eq.~\eqref{eq:exposure_based_constraint}.

In contrast, the uniform ranking policy $\piunif$ provides the following impact on the $n$-th item:
\begin{align*}
    \imp_n (X_{*,n,*}^{\piunif}) = \sum_u \frac{n-u+1}{n^2} \frac{\te}{n} = \frac{n+1}{2n^2} \te
\end{align*}
Therefore,
\begin{align*}
    \frac{\imp_n  (X_{*,n,*}^{\piexpo})}{\imp_n (X_{*,n,*}^{\piunif})} \le \frac{4n}{(n+1)^2} = \calO ( n^{-1} ).
\end{align*}

In addition, $e_{max} := \max_k e(k)$ needs to be larger than $\expo_1(X^{\piexpo})/n$ so that the exposure-fair constraint is feasible. Given that $\piexpo$ maximizes the utility to the users under the exposure-fair constraint and allocates exposure of the most relevant users preferentially to the 1st item, we have
\begin{align}
    \imp_n (X_{*,1,*}^{\piexpo}) & \ge  \sum_u \frac{n-u+1}{n^2} \frac{\expo_1(X^{\piexpo})}{n} = \frac{\te}{n}
    \label{eq:a_4}
\end{align}
Eq.~\eqref{eq:a_4} provides the lowest possible utility of the $n$-th item under $X_{*,1,*}^{\piexpo}$ when $e_{max}$ is the lowest possible with $$e_{max} = \frac{\expo_1(X^{\piexpo})}{n} = \frac{2\te}{n+1}.$$

Using Eqs~\eqref{eq:a_3} and \eqref{eq:a_4}, we have
\begin{align*}
    \frac{\max_{j\in \calI} \imp_n (X_{*,j,*}^{\piexpo})}{\imp_n (X_{*,n,*}^{\piexpo})} = \frac{\imp_n (X_{*,1,*}^{\piexpo})}{\imp_n (X_{*,n,*}^{\piexpo})} \ge \frac{n+1}{2}
\end{align*}

\vspace{2mm}
\paragraph{Reduction used in Theorem~\ref{thm:nsw}}
If $K=1$, our problem is reduced to the classical fair division of divisible goods with additive utility. Thus, the NSW-maximizing solution is known to satisfy Pareto optimality, envy-freeness, and dominance over $\piunif$ (which corresponds to PFS)~\citep{varian1974equity}. For a general case with $K>1$, our problem is reduced to fair division with \textit{constrained groups} described in Appendix A of~\citet{kroer2019scalable}. Thus, our result follows from a version of their Theorem 1.

In more detail, fair division with \textit{constrained groups} is an instance of fair division with linear utility function where we have a set $\mathcal{P}$ of constraint groups, which is a partitioning of resources. Each $P \in \mathcal{P}$ is a subset of the original resources that should be allocated to the agents. It is assumed that each agent can be assigned 1 unit of resources among those in each group $P$.
This problem of fair division with constrained groups is reduced to our problem of fair ranking when we regard agents as items, resources as positions in a ranking, and a constraint group as a set of positions in a ranking that is shown to each user $u$. In the ranking problem, a set of positions ($P_u \in \mathcal{P}$) that is associated with a single user $u$ can be allocated to each item in total quantity of $1$ (i.e., a constraint described as $\sumK \allocation = 1, \, \forall (u, i)$) as in fair division with constraint group. 

Under this reduction and the assumption that there exists at least $K+1$ $\epsilon$-twins for each item, Theorem 1 in Appendix A.2 of~\citet{kroer2019scalable} suggests that envy and the proportional share gap (which is $\imp_i(X^{\piunif}_{*,i,*})  - \imp_i(X^{\pinsw}_{*,i,*})$ in our problem) is upper bounded by $\epsilon (c_1 + n/(m c_2))$ for some positive constants $c_1,c_2>0$ where the notations in the original paper, $\sum_i B_i$ (sum of agent budgets) and $|\mathcal{K}|$ (number of constraint groups) can be regarded as $n$ (number of items) and $m$ (number of users) in our setup, respectively.

\begin{table}[t]
\caption{Statistics of the real-world datasets after preprocessing and the relevance prediction accuracy on the test set} \label{tab:data_stats}
\centering
\scalebox{0.9}{\begin{tabular}{c|cc}
\toprule
  & Delicious & Wiki10-31k \\ \midrule
Training Data Size & 9,123 & 7,639  \\ 
Test Data Size ($|\calU|$) & 1,014 & 849  \\ 
Number of Labels ($|\calI|$) & 100 & 100 \\ \midrule
AUC (on test set) & 0.81 &0.87 \\ 
LogLoss (on test set) & 0.31 & 0.18 \\ 
\bottomrule
\end{tabular}}
\end{table}
\begin{figure*}[h]
\includegraphics[width=16.5cm]{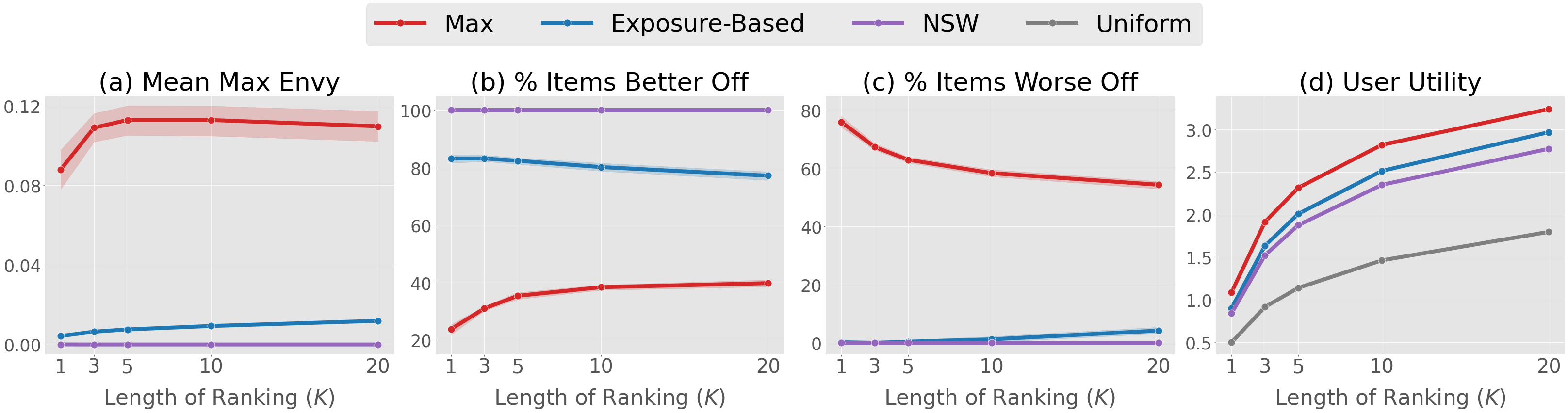}
\caption{Fairness and user utility on synthetic data with varying lengths of ranking.}
\label{fig:k}
\end{figure*}

\begin{figure*}[h]
\includegraphics[width=16.5cm]{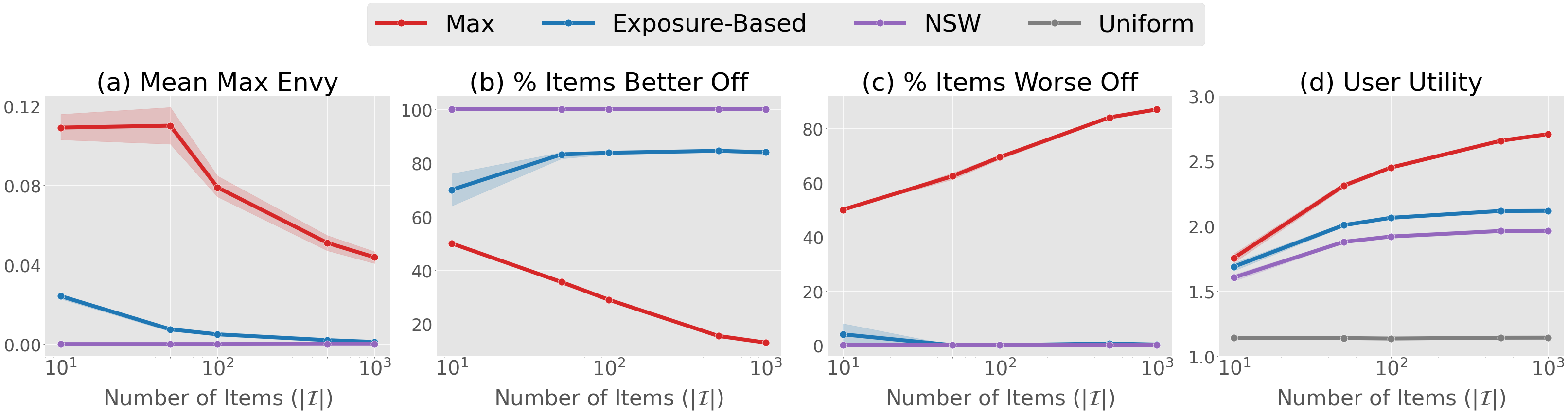}
\caption{Fairness and user utility on synthetic data with varying numbers of items.}
\label{fig:n_items}
\end{figure*}

\begin{figure*}[h]
\includegraphics[width=16.5cm]{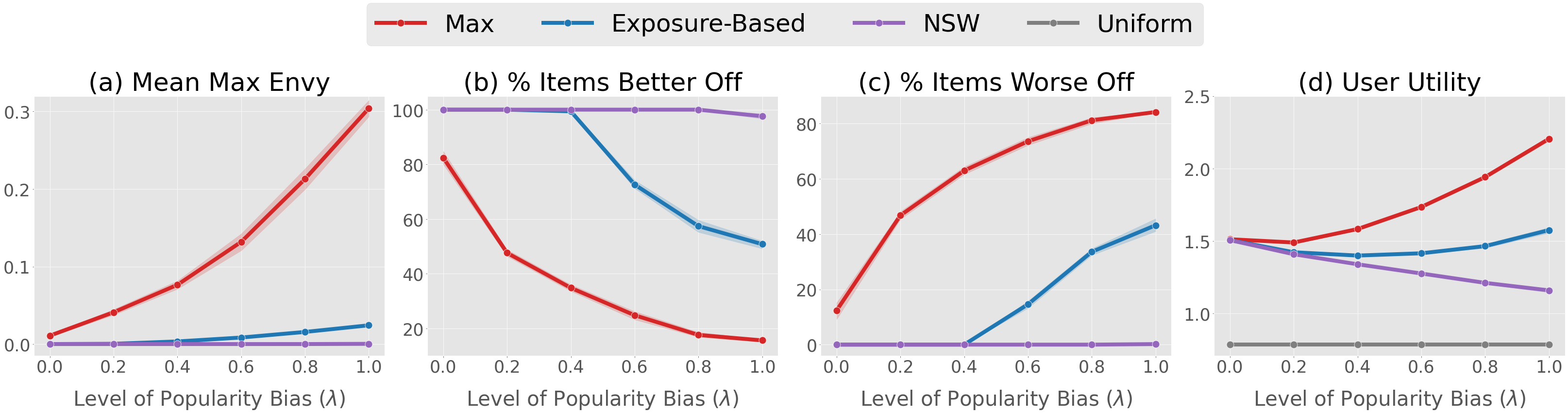}
\caption{Fairness and user utility with different popularity patterns and with exponential examination function.}
\label{fig:exponential}
\end{figure*}

\begin{figure*}[h]
\includegraphics[width=16.5cm]{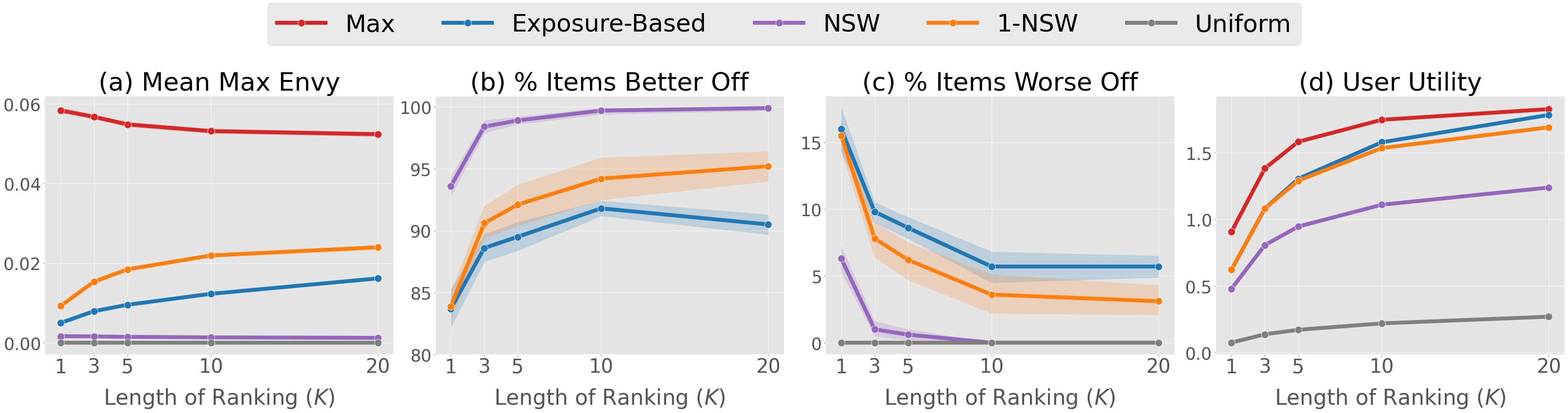}
\caption{Fairness and user utility on the Wiki10-31K dataset with varying lengths of ranking.}
\label{fig:wiki}
\end{figure*}

\section{Experiment Details and Results} \label{app:experiment_detail}

\subsection{Synthetic Data}

In this section, we employ the same default setting as in Section~\ref{sec:synthetic} and vary some additional configurations.

\subsubsection{How do the ranking policies perform with different lengths of ranking?}
Here, we validate how the policies perform as the length of ranking ($K$) varies. Figure~\ref{fig:k} reports the results with $K \in \{1,3,5,10,20\}$. First, we can see that all policies improve user utility with increasing $K$ in Figure~\ref{fig:k} (d), as more items contribute to the user utility.
As for the fairness measures, $\pinsw$ remains envy-free and improves impact on all items over 10\% from $\piunif$, while $\piexpo$ produces a larger amount of envy and a less equitable impact distribution with increasing $K$.

\subsubsection{How do the ranking policies perform with different market sizes?}
Next, we vary the size of the market. Figure~\ref{fig:n_items} shows how the fairness measures and user utility of the ranking policies change with varying numbers of items and a fixed number of users. We observe that $\pimax$ produces a larger amount of envy and a less equitable impact distribution with the growing market size. In contrast, $\piexpo$ slightly improves both the fairness measures and user utility, but $\pinsw$ is much fairer for all cases, always being almost envy-free and dominating $\piunif$ in terms of impact distribution. $\piexpo$ does not achieve these desiderata for all market sizes.

\subsubsection{How does the examination function influence the relative performance?}
Finally, we evaluate whether a different examination function results in a different conclusion. We use the inverse function $e(k) := \mathbb{I} \{k \le K\}/k$ in the main text, but there are some other types of examination functions we can consider~\citep{su2021optimizing}. Here we use the exponential examination function defined as $e(k):=\mathbb{I} \{k \le K\}/\exp(k-1)$, which has a steeper drop off in probability than the inverse function, and as a result, assumes that users are likely to only see the top few items. Figure~\ref{fig:exponential} shows the item fairness measures and user utility with varying popularity patterns ($\lambda \in \{0.0,0.2,\ldots,1.0\}$) and with the exponential examination function. In general, we observe trends similar to Figure~\ref{fig:pop} (which is obtained with the inverse examination function) for all metrics. Because we use a steeper examination function, the user utility decreases for all policies compared to Figure~\ref{fig:pop}, but the relative fairness and utility remain similar.

\subsection{Real-World Data}
Table~\ref{tab:data_stats} summarizes some statistics of the datasets after applying the preprocessing described in the main text. Figure~\ref{fig:wiki} shows the fairness measures and user utility of $\pimax$, $\pinsw$, $\pinswone$, $\piexpo$, and $\piunif$ with varying $K$ on Wiki10-31K. We observe that $\pinsw$ is almost envy-free and achieves an equitable impact distribution for all values of $K$. In addition, $\pinswone$ achieves similar levels of user utility as $\piexpo$, while leading to a fairer impact distribution.

\end{document}